\documentstyle[prl,floats,aps,twocolumn,epsf,graphicx]{revtex}
\begin{document}
\twocolumn[\hsize\textwidth\columnwidth\hsize\csname
@twocolumnfalse\endcsname

\title{Self-Segregation vs. Clustering in the Evolutionary Minority Game}
\author{Shahar Hod$^1$ and Ehud Nakar$^2$}
\address{$^1$Department of Condensed Matter Physics, Weizmann Institute, Rehovot 76100, Israel}
\address{}
\address{$^2$The Racah Institute of Physics, The
Hebrew University, Jerusalem 91904, Israel}
\date{\today}
\maketitle

\begin{abstract}

\ \ \ Complex adaptive systems have been the subject of much recent attention. It is 
by now well-established that members (`agents') tend to self-segregate into opposing 
groups characterized by extreme behavior. However, while different social and biological 
systems manifest different payoffs, the study of such adaptive systems has mostly been 
restricted to simple situations in which the prize-to-fine ratio, $R$, equals unity. In this Letter 
we explore the dynamics of evolving populations with various different values of the ratio $R$, 
and demonstrate that extreme behavior is in fact {\it not} a generic feature of adaptive systems. 
In particular, we show that ``confusion'' and ``indecisiveness'' take over in times of depression, in 
which case cautious agents perform better than extreme ones.
\end{abstract}
\bigskip

]

A problem of central importance 
in social, biological and economic sciences is that of an evolving population in which 
individual agents adapt their behavior according to past experience, without direct 
interaction between different members.
Of particular interest are situations in which members (usually referred to as `agents') 
compete for a limited resource, 
or to be in a minority (see e.g., \cite{John1} and references therein.) 
In financial markets for instance, more buyers than sellers implies higher prices, 
and it is therefore better for a trader to be in a minority group of sellers. 
Predators foraging for food will do better if they hunt in areas with fewer competitors. 
Rush-hour drivers, facing the choice between two alternative routes, wish to choose 
the route containing the minority of traffic \cite{HubLuk}. 

Considerable progress in the theoretical understanding of such systems has been gained 
by studying the simple, yet realistic model of the minority game (MG) \cite{ChaZha}, 
and its evolutionary version (EMG) \cite{John1} (see also 
\cite{DhRo,Cev,BurCev,LoHuJo,HuLoJo,HaJeJoHu,BuCePe,LoLiHuJo,LiVaSa,SaMaRi} 
and references therein). 
The EMG consists of an odd number of 
$N$ agents repeatedly choosing whether to 
be in room `0' (e.g., choosing to sell an asset or taking route A) or in room `1' 
(e.g., choosing to buy an asset or taking route B). 
At the end of each turn, agents belonging to the smaller group (the minority) are 
the winners, each of them gains $1$ point (the `prize'), 
while agents belonging to the majority room lose $1$ point (the `fine'). 
The agents have a common `memory' look-up table, containing the outcomes of $m$ 
recent occurrences (the particular value of $m$ is of no importance \cite{John1}). 
Faced with a given bit string of recent $m$ occurrences, each agent chooses 
the outcome in the memory with probability $p$, known as the agent's ``gene'' value 
(and the opposite alternative with 
probability $1-p$). If an agent score falls below some value $d$, then its strategy 
(i.e., its gene value) is modified. In other words, each agent tries to learn from his 
past mistakes, and to adjust his strategy in order to survive.

A remarkable conclusion deduced from the EMG \cite{John1} 
is that, a population of 
competing agents tends to self-segregate into opposing 
groups characterized by extreme behavior. 
It was realized that in order to flourish in such situations, an agent 
should behave in an extreme way ($p=0$ or $p=1$) \cite {John1}. 

It should be emphasized, however, that previous analyses were restricted to the simple 
case in which the prize-to-fine ratio, $R$, was assumed to be equal unity. 
On the other hand, in many real life situations this ratio may take a variety of 
different values. In the extreme situation, the fine (e.g., a temporary worker 
getting fired of work 
after being late to the office due to a traffic jam, or a predator being starved to 
death while unsuccessfully trying to hunt in an area with many competitors) 
may be larger than the prize (a day's payment or a successful hunt which guarantees 
food for few days, respectively). Another example is that of a trader in a 
financial market which is under depression. In such circumstances, the trader usually lose more money 
in a bad deal than he gains in a successful one (due to overall reduction in market's value). 
In addition, in many cases there are taxes on profits, implying a reduction in the value of $R$.

Moreover, we know from real life situations that extreme agents not always perform better 
than cautious ones. In particular, our daily experience indicates that in difficult situations 
(e.g., when the prize-to-fine ratio is low) human people tend to be 
confused and indecisive. In fact, in such circumstances 
they usually seek to do the {\it same} (rather than the opposite) as the majority. 

Thus, of great interest for real social and biological systems are situations in which the 
prize-to-fine ratio is smaller (or larger) than unity. 
The aim of the present Letter is to explore the dynamics of evolving populations with 
various different external conditions (i.e., different values of the ratio $R$). Of main importance 
is the identification of the strategies that perform best in a particular situation.

\begin{figure}[tbh]
\centerline{\epsfxsize=9cm \epsfbox{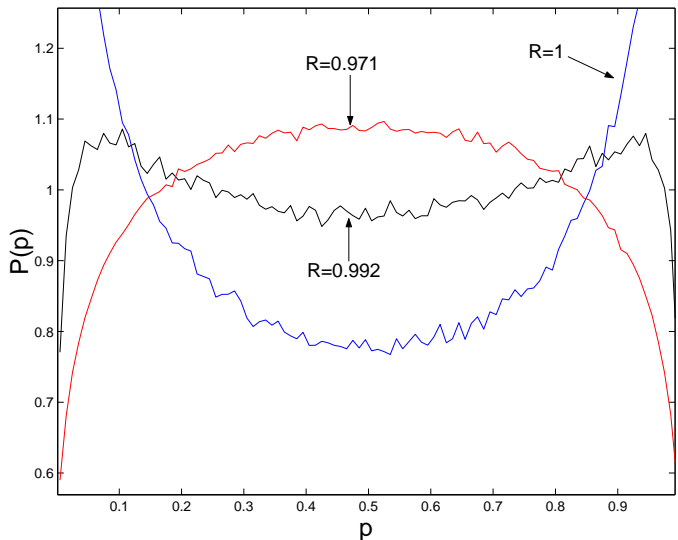}} 
\caption{The strategy distribution $P(p)$ for different values of the 
prize-to-fine ratio: $R=0.971, R=0.992$, and $R=1$. 
The results are for $N=10001$ agents, $d=-4$. Each point represents 
an average value over 10 runs and 100000 time steps per run.}
\label{Fig1}
\end{figure}

Figure \ref{Fig1} displays the long-time frequency distribution $P(p)$ of the agents (the lifespan, $L(p)$, 
defined as the average length of time a strategy $p$ survives between modifications, has a similar 
behavior). We find three qualitatively different populations, depending on the precise value of the 
prize-to-fine ratio parameter, $R$.
For $R > R^{(1)}_c$ (this includes the case studied so far in the literature, $R=1$. 
The value of $R^{(1)}_c$ depends on the number of agents and the parameter $d$) the 
distribution becomes peaked around $p=0$ and $p=1$ -- the population will self-segregate 
(this corresponds to always or never following what happened last time). To 
flourish in such a population, an agent should behave in an {\it extreme} way \cite{John1}. 
On the other hand, for $R < R^{(2)}_c$ (poor conditions, in which the fine is larger than 
the reward) the population tends to crowd around $p={1 \over 2}$. 
This corresponds to ``confused'' and ``indecisive'' agents. There is 
also an intermediate phase [for $R^{(2)}_c < R < R^{(1)}_c$], in which the population tends to 
form an M-shaped distribution, peaked around some finite $p_0$ and its counterpart $1-p_0$ (with 
the absolute minimas of the distribution located at $p=0$ and $p=1$).

\begin{figure}[tbh]
\centerline{\epsfxsize=9cm \epsfbox{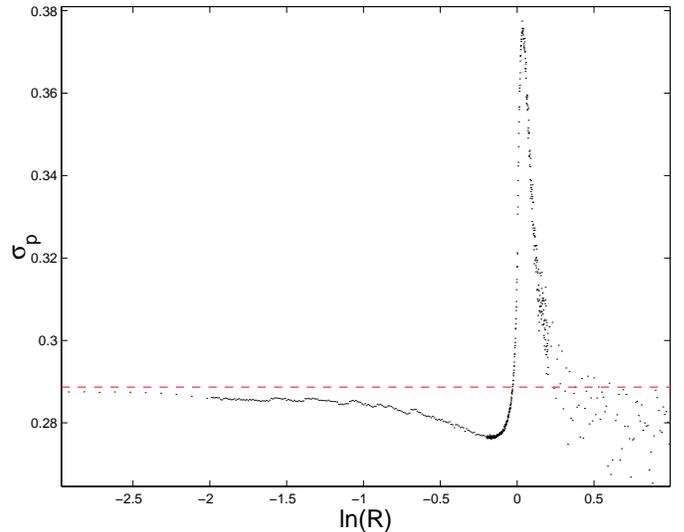}} 
\caption{The root-mean-square separation (rms) of the strategies as a function of the prize-to-fine 
ratio, $R$. Horizontal line represents the rms separation for a uniform $P(p)$ distribution. 
$N=1001,\  d=-4$. Each point represents an average value over 10 runs and 10000 time steps per run.}
\label{Fig2}
\end{figure}

An important feature of the original EMG (for the $R=1$ case \cite{John1}) is that 
the root-mean-square (rms) 
separation of the strategies is {\it higher} than the corresponding value for uniform $P(p)$. This indicates the 
desire of agents to do the {\it opposite} of the majority \cite{John1}. Figure \ref{Fig2} shows the rms separation 
of the population as a function of the prize-to-fine ratio, $R$. Remarkably, we find that for small values of
$R$ the rms is in fact {\it smaller} than that obtained for a uniform $P(p)$ distribution. We therefore 
conclude that in times of difficulties agents desire to do the {\it same} (rather than the opposite) as the majority. 
This is exactly the type of behavior we anticipated in the introduction based on daily experience.

\begin{figure}[tbh]
\centerline{\epsfxsize=9cm \epsfbox{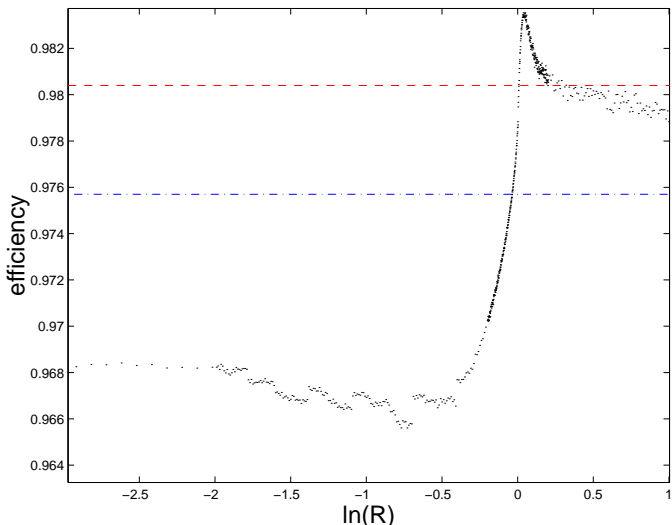}} 
\caption{The (scaled) efficiency $E$ of the system as a function of the prize-to-fine 
ratio, $R$. Horizontal lines represents the efficiency for uniform $P(p)$ distribution (dashed) and 
a coin-tossing situation (dashed-dotted). The parameters are the same as in Fig. 2.}
\label{Fig3}
\end{figure}

In the original EMG \cite{John1} it was found that the dynamics of the system leads to 
situations in which the size of the minority group is maximized, indicating that the efficiency 
of the system is maximized. 
The (scaled) efficiency of the system is defined as the number of agents in the 
minority room, divided by the maximal possible size of the 
minority group, $(N-1)/2$. 
Figure \ref{Fig3} displays the system's efficiency as a function of the ratio $R$. We also display the 
efficiency for agents guessing {\it randomly} between room `0' and room `1', and for a {\it uniform} 
distribution of agents. As previously found, 
there is a range of $R$ (which includes the previously studied case, $R=1$ \cite{John1}) 
in which the efficiency of the system is {\it better} than the random case. 
However, for small values of the prize-to-fine ratio, 
the efficiency of the system is remarkably {\it lower} than that obtained for agents choosing via 
independent coin-tosses. Thus, considering the efficiency of the system as a whole, 
the agents would be better off not adapting their 
strategies because they are doing {\it worse} than just guessing at random.

Note that an optimum utilization of the resources is obtained at 
some $R_{max}>1$ (with $R_{max}-1 \ll 1$). 
This implies that an evolving population requires a small positive feedback in order to exploit 
its resources in an optimal way. On the other hand, a wealthy society has an efficiency which is 
worse than that of a uniform $P(p)$ distribution (this occurs for prize-to-fine ratios which are 
too large). This reflects the fact that in a ``spoiling'' environment, the agents has no real 
motivation to evolve (they have a long lifespan even without exploiting their resources in an 
optimal way). 

\begin{figure}[tbh]
\centerline{\epsfxsize=9cm \epsfbox{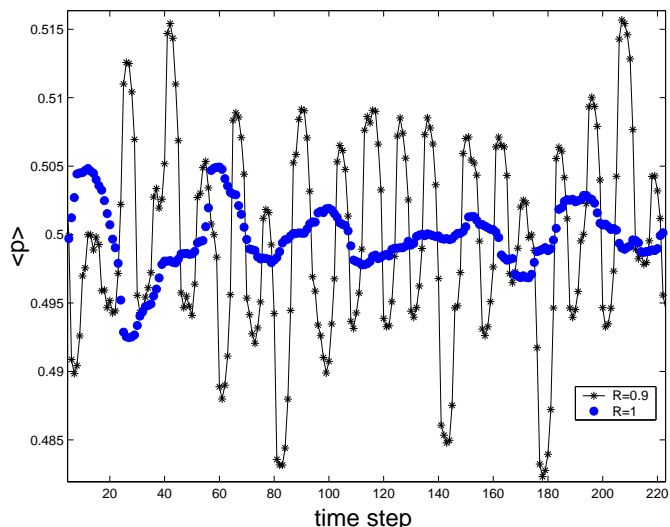}} 
\caption{Time evolution of the average gene-value, $<$$p$$>$, for different values of the 
prize-to-fine ratio: $R=0.9$ and $R=1$. The parameters are the same as in Fig. 1.}
\label{Fig4}
\end{figure}

In previous studies (of the $R=1$ case) it has been established that the evolving population enters into 
a {\it stationary} phase, in which case the $P(p)$ distribution remains essentially 
constant in time \cite{LoHuJo,DhRo}. 
In Figure \ref{Fig4} we display the time-dependence of the average gene value, $<$$p$$>$, for different values of the 
prize-to-fine ratio, $R$. The distribution $P(p)$ oscillates around $p={1 \over 2}$. 
The smaller is the value of $R$ the larger are the amplitude and the frequency of the oscillations. 
Thus, we conclude that a population which evolves in a tough environment 
never establishes a steady state distribution. Agents are constantly changing 
their strategies, trying to survive. By doing so they create global currents in the gene space. 

We now provide some analytical analysis of the problem, a generalization of 
the one presented in \cite{John1} for 
arbitrary values of the prize-to-fine ratio, $R$. The simplest example of our 
system contains $N=3$ agents, and three discrete gene values $p=0,\ {1 \over 2},\ 1$. We 
consider configurations for which the average gene value lies between ${1 \over 3}$ and ${2 \over 3}$, 
a reminiscent of the fact that $<$$p$$>$ displays only mild oscillations around $p={1 \over 2}$. 
To obtain the average $P(p)$ distribution we weight the various 
configurations according to the average points awarded per agent in each of the configuration \cite{Note1}. 
The analytical results are given in Table \ref{Tab1} [Note that $P(0)=P(1)$.] We find that this 
simplified toy-model provides a fairly good qualitative description of the complex system. 
In particular, it follows that the population self-segregates for prize-to-fine ratios larger than 
$R_c={5 \over 7}$, while for $R<R_c$ the agents tend to cluster around $p={1 \over 2}$. 
In addition, the efficiency of the system is maximal at {\it intermediate} values of the prize-to-fine ratio, 
while poor ($R<2$) and 
wealthy ($R>3$) populations display a lower efficiency.

\begin{table}
\caption{Distribution of strategies and efficiency of a three agents system.}
\label{Tab1}
\begin{tabular}{lll}
$R$ &$P(0):P({1 \over 2})$ & $efficiency$\\
\tableline
$R<2$ & ${{19R-53} \over {26R-58}}$ & ${{57R-150} \over {64R-164}}$ \\
$2 \le R \le 3$& $2.5$ & $1$ \\
$R>3$& ${{23R-49} \over {23R-61}}$ & ${{85R-191} \over {92R-212}}$ \\
\end{tabular}
\end{table}

In summary, we have explored the evolution of complex adaptive systems with an arbitrary value 
of the prize-to-fine ratio, $R$. 
The main results and their implications are as follows:

(1) It has been widely accepted that {\it self-segregation} is 
a generic characteristic of an evolving population of competing agents. This belief was based on studies 
of the $R=1$ case. Our analysis, however, turns over this point of view. 
In particular, in times of difficulties agents tend to {\it cluster} around $p={1 \over 2}$ -- 
cautious agents perform better (live longer) than extreme ones. Stated in a more pictorial way, 
confusion and indecisiveness take over at tough times. 

(2) In previous analyses it was found that agents desire to do the opposite of the majority \cite{John1}. We have 
shown that this property is in fact not a generic one. 
In particular, in a tough environment agents try to do the {\it same} as the majority 
[The rms separation of strategies is in fact {\it smaller} than that obtained for a uniform $P(p)$ distribution.]

(3) For small values of the prize-to-fine ratio (poor external conditions) 
the efficiency of the system is well below the efficiency achieved by random agents (ones who choose via 
independent coin-tosses). It seems that ``panic'' and ``confusion'' (clustering around $p={1 \over 2})$ 
prevent the agents from achieving a reasonable utilization of resources. Similarly, a wealthy population, 
for which there is no real motivation for adaptation, displays a poor efficiency. 
On the other hand, an evolving population achieves an 
optimum utilization of its resources when it receives a ({\it small}) positive external reinforcement 
(that is, for $0 < R_{max}-1 \ll 1$). 

(4) The gene-distribution, $P(p)$, displays temporal-oscillations around $p={1 \over 2}$. 
The smaller is the value of the prize-to-fine ratio, the farther is the system from a 
steady-state distribution. This in particular implies that the steady-state assumption used to 
analyze the EMG (in the $R=1$ case) \cite{LoHuJo} is no longer valid for smaller values of $R$.

\bigskip
\noindent
{\bf ACKNOWLEDGMENTS}
\bigskip

SH thanks Mordehai Milgrom for his kind assistance.
The research of SH was supported by grant 159/99-3 from the Israel Science Foundation.

\end{document}